\documentstyle [preprint,prl,aps]{revtex}

\begin{document}
\draft

\title{ Structural and Electronic Properties of a Carbon
Nanotorus: Effects of Non-Local Vs Local Deformations}
\author{Lei Liu, C. S. Jayanthi, and S. Y. Wu}
\address {Department of Physics, University of Louisville,
Louisville, Kentucky 40292}
\maketitle

\begin{abstract}

The bending of a carbon nanotube is studied by considering
the structural evolution of a carbon nanotorus  
from elastic deformation to the onset of the kinks and 
eventually to the collapse of the walls of the nanotorus.  
The changes in the electronic properties due to {\it non-local}
deformation are contrasted with those due to {\it local}
deformation to bring out the subtle issue underlying the reason
why there is only a relatively small reduction in the electrical
conductance in the former case even at large bending angles while
there is a dramatic reduction in the conductance in the latter
case at relatively small bending angles.

\end{abstract}


There has been some focused interest in the last several years on the 
interplay between the mechanical deformation and the electrical 
properties of single-wall nanotubes (SWNT). 
Both experimental\cite{Dekker98,Paulson99} and 
theoretical\cite{Crespi97,Kane97,Nardelli98,Avouris99} 
studies seem to indicate that bending of the nanotubes 
induces only a small conductance change even at
relatively large bending angles (up to $\alpha=45^o$
\cite{Avouris99}, where $\alpha$ is the angle between the 
direction tangential to the end of the tube and the unbent
axis) unless the nanotubes fracture or the metal-tube contacts 
are perturbed \cite{Paulson99}. 
Bending involved in these studies is caused by a mechanical deformation 
under forced confinement of the ends of the tube\cite{Bernholc96}. 
In this situation, the 
deformation induced by the bending can be viewed as a non-local
deformation. Very recently, we carried out an in-depth study, both 
theoretically and experimentally, of the effects of a local mechanical 
deformation of a metallic SWNT induced by the pushing action of the 
tip of an atomic force microscope (AFM) on the electrical properties 
of the tube\cite{Liu00,Dai00}. We observed an {\it unexpected} 
two orders of magnitude reduction in conductance at a 
relatively small bending angle ($\alpha=13^o$)\cite{Dai00}. 
Our theoretical study confirmed that this dramatic change in conductance 
is due to a reversible transition from $sp^2$ to $sp^3$ bonding configurations 
in the bending region adjacent to the tip\cite{Liu00}. Thus, it is the local 
deformation induced by the pushing action of the tip that is responsible 
for the two orders of magnitude change in conductance.  This
raises the question why there is only a relatively small change in conductance even at 
large bending angles under a non-local deformation? In this work, we address 
this question by investigating the mechanical deformation of a carbon
nanotorus obtained by connecting the two ends of a SWNT into a
ring.

We have chosen a carbon nanotorus to study the effects
of bending because in this case one can determine readily the radius
of curvature, the logical quantity to characterize the degree of
bending, at the location of bending than in the case of a bent SWNT.
In previous theoretical studies\cite{Bernholc96,Nardelli99,Avouris98}, 
the bending of the SWNT was modeled 
by holding the ends of the SWNT at positions defining the angle of 
bending. This initial configuration was then allowed to relax to its 
equilibrium configuration while the two ends were kept at the 
initial positions. Thus, there could be some ambiguity in 
relating the bending angle set initially with the deformation in 
the bending region. On the other hand, the radius of curvature of 
a carbon nanotorus is well defined. Hence, a nanotorus is ideally 
suited for studying the effects of non-local deformations and
can,  in fact, lead to an understanding of the  
interplay between the mechanical deformation and electronic properties 
of SWNTs. Furthermore, recent experimental fabrication of rings from 
SWNTs\cite{Avouris99a} and subsequent measurement of 
magneto-resistance\cite{Avouris00} from some of these rings underscore 
the importance of understanding the interplay 
between the mechanical deformation and the electronic properties
of a nanotorus in its own right.

To study how non-local deformations affect the structure 
of a nanotube, we consider a metallic (5,5) nanotube bent into a 
circular form with its two ends connected to form a
nanotorus of a certain radius (R). The initial configuration is
then relaxed to its equilibrium configuration. 
To obtain a nanotorus which is uniformly and elastically
deformed, one must start with a nanotube of sufficiently long
length. Because of the size of the 
system under consideration, we used the order-N non-orthogonal
tight-binding molecular dynamics (O(N)/NOTB-MD) \cite{Wu98} with the NOTB 
Hamiltonian as given in Ref.\cite{Menon93}. We established that a 
relaxed nanotorus containing 2000 atoms will 
indeed be under a uniform and elastic deformation. This was
accomplished by first obtaining the relaxed configuration of
a straight 2000-atom (5,5) nanotube (relaxed radius=3.42 {\AA}),
next bending it into a circular form, and finally relaxing it into its 
equilibrium configuration. Starting from this configuration, we studied the 
effects of increasing 
deformation by reducing the radius of the nanotorus in a quasi-continuous 
step-by-step manner.
Our aim is to effectively  model a continuous bending process. 
Thus, at each step, we removed a small 
segment containing two circular sections, each with 10 atoms, from
the nanotorus. The maximum change in the bond length with the removal
of 20 atoms was less than 0.01{\AA} when the two ends of the broken nanotorus 
were elastically reconnected. This change represents a small 
perturbation in the strain going from one step to the next. The 
configuration so obtained at each step was then relaxed to its 
equilibrium configuration using the O(N)/NOTB-MD scheme. The 
procedure was continued from the stage of elastic deformation, to 
the stage of the development of the kinks, and eventually to the 
stage of topological change in the structure in the bending region.

We highlight the "continuous" process of bending by four configurations 
shown in Fig.1. Figures 1(a)-1(d) give the equilibrium configurations 
obtained by the MD procedure outlined above for nanotori containing 
1000 (R=19.6{\AA}), 900 (R=17.6{\AA}), 800 (R=15.7{\AA}), and 680 (R=13.3{\AA}) 
atoms respectively. An examination of the structures shown in Fig. 1 
reveals the following features. From Fig. 1(a), it can be seen that 
the nanotorus of R=19.6{\AA} is under a uniform elastic distortion, 
with a simple stretching on the outer side and simple compression on 
the inner side. It gives an example of a SWNT under elastic deformation 
and indicates that the (5,5) SWNT is still only elastically deformed 
at a radius of curvature (R) of 19.6{\AA} under a uniform deformation. As  
R is reduced to 17.6{\AA} with a corresponding increase in the 
mechanical deformation, the structural distortion is characterized by 
the appearance of 16 almost uniformly distributed kinks along the inner 
(compression) side of the nanotorus (see Fig. 1(b)). The appearance of 
the kinks is attributable to the release of the strain energy at a 
critical radius of curvature of about 17.6{\AA} in the bending region. 
Further reduction of R with the accompanying increase in the 
distortion brings about the reduction of the number of kinks
while enhancing the release of the strain energy at the kinks as exemplified 
in Fig. 1(c). Finally when R reaches 13.3{\AA}, the collapsing of the 
inner and outer walls of the nanotorus occurs (see Fig. 1(d)), signaling a 
topological change in the structure in the region of collapse at a critical 
radius of curvature of about 13.3{\AA}. During the simulation of the
"continuous" bending, extreme care was exercised in monitoring the onset of the 
fundamental structural change. However, due to the discrete nature of the 
removal of small segment from the nanotorus, there is a built-in degree 
of uncertainty in pinning down precisely the radius of curvature when a 
certain structural change occurs. Hence the numbers given above as 
critical radii of curvature for various structural changes are approximate.

To explore in more detail the structural changes induced by a
non-local deformation, we calculated the average strain energy per atom over 
each section (containing 10 atoms) of the relaxed nanotorus. The
angular location of the section is denoted by ($\theta$) with
$\theta$=0 along the positive horizontal direction and increasing
in the clockwise direction. The total energy of the 
nanotorus can be expressed as the sum of the site energy of atoms 
in the nanotorus 
within the framework of a NOTB Hamiltonian \cite{Wu99}. The strain energy of 
an atom at a given site in the relaxed (5,5) nanotorus is defined as 
the energy difference between the energy of the atom at that site 
and the energy per atom for the relaxed ideal straight (5,5) nanotube. The 
average strain energy per atom for a given section is simply the average 
of the strain energy for the atoms in that section. In Fig. 2, the 
average strain energy per atom for the four nanotori is plotted against 
$\theta$. For the nanotorus of R=19.6 {\AA}, the strain 
energy is small and uniform along the entire torus, indicating that 
the torus is under simple elastic deformation. When R is reduced 
to 17.6{\AA}, the average strain energy along the torus is 
increased substantially and 16 local minima appear in the plot at the 
angles where the kinks are located, a clear indication of the release 
of the strain energy at the kinks. When R is reduced to 
15.7{\AA} (800 atoms), there is only a slight increase in the average strain 
energy at locations near the kinks with enhanced local 
minima. However, when R reaches 13.3{\AA}, 
pronounced local minima appear at locations precisely matching where 
collapsing of the walls occurs. These enhanced local minima actually 
accompany a substantial increase in the average strain energy per atom 
along the torus. A side-by-side comparison of Fig. 2 and Fig. 3, where 
the average bond length is plotted vs $\theta$ , is even more illuminating. 
While the average bond length along the torus is somewhat longer than 
the bond length of the ideal straight (5,5) nanotube for the three 
nanotori with the radius of 19.6 {\AA}, 17.6 {\AA}, and 15.7{\AA} respectively, the 
average bond length for the nanotorus with the radius 13.3{\AA} at the 
locations of wall-collapsing is 1.53{\AA}, substantially longer than the 
bond length of 1.42{\AA} for the $sp^2$ bonding configuration in the ideal straight 
(5,5) nanotube and in fact close to the bond length of 
1.54{\AA} for the $sp^3$ bonding configuration in diamond. This is the 
signature of the topological change in the structure for the 680-atom 
nanotorus in the region of wall-collapsing. Further confirmation of 
this picture can be seen from Fig. 4 where the average coordination 
number is plotted vs $\theta$ . It can be seen that the average coordination 
number at the locations of wall-collapsing for the nanotorus of radius 
13.3{\AA} is about 3.6, close to the 4-coordinated $sp^3$ configuration, 
while that along the circumference of the first three nanotori is 
close to 3, the coordination number for the $sp^2$ configuration.  Thus, 
our investigation has succinctly established the effects on the structure 
of a SWNT due to a non-local deformation, namely, from a simple 
elastic distortion for large radii of curvature, to the appearance of 
kinks at a critical radius of curvature of about 17.6{\AA}, followed by 
the collapse of the walls involving the transition from the $sp^2$ 
configuration to the $sp^3$ configuration at a critical radius of 
curvature of about 13.3{\AA} for the (5,5) SWNT. Within the framework 
of the nanotorus, one may set up a benchmark for comparison with 
previous studies. Specifically, referring to Fig. 1(b), since there 
are 16 uniformly distributed kinks, one may estimate the critical angle 
for the appearance of the kink under a non-local deformation as 
$\alpha_K= \frac {1}{2}(\frac {2 {\pi}}{16})\simeq 11^o$. 
Similarly, from Figs. 2(d), 3(d), and 4(d), corresponding to the 680-atom case, 
one may surmise that there should be 7 equivalent pronounced peaks or 
valleys at the locations of wall-collapsing along the circumference. 
Hence, the critical bending angle for the transition from $sp^2$- to 
$sp^3$-bonding configuration under a non-local deformation can be estimated 
as $\alpha_C= \frac {1}{2} (\frac {2\pi}{7}) \simeq 26^o$. But, care must be exercised 
when using the bending angle because 
of the ambiguity in relating the preset bending angle to the degree 
of bending in the bending region. Unless the arc length defining the 
bending region is also known, the preset bending angle alone can not 
uniquely characterize the degree of bending.

To understand the interplay between the mechanical deformation and the 
electronic properties, we calculated the average $\sigma$ bond charge and the 
average $\pi$ bond charge per atom following the procedures given in Refs. 
\cite{Wu99,Liu00}. The results are shown in Figs. 5 and 6,
respectively. It can be seen that,
for the first three nanotori, there is only some slight reduction of 
both the $\sigma$ and the $\pi$ bond charges along the circumference associated 
with the bond stretching as the nanotorus is under increasing strain. 
Thus, there is no basic change in the nature of electronic structure and 
hence no significant change in conductance is expected for a SWNT under 
a non-local deformation up to a radius of curvature of about 13.3{\AA}.  
However when the strain reaches a critical value such that 
wall-collapsing occurs ($R\le$13.3{\AA}), there is a substantial increase 
in the $\sigma$ bond charge and a pronounced decrease in the $\pi$ bond charge at the 
locations where the walls collapse, a reflection of the transition 
from the $sp^2$ bonding configuration to the $sp^3$ bonding
configuration. (It should be noted that this is precisely the mechanism 
which comes into play for a SWNT under a local deformation at relatively 
small bending angles \cite{Liu00,Dai00}). The $\pi$ electrons are delocalized and 
mainly responsible for conduction while the $\sigma$ electrons are localized.  
Therefore one might expect that 
the dramatic reduction of the $\pi$ bond charge and substantial increase 
in the $\sigma$ bond charge will lead to a substantial reduction in conductance 
for a nonlocal deformation for R$\le$ 13.3{\AA} (i.e. $\alpha$ $\ge$26$^o$).
We, therefore, calculated the electrical
conductance corresponding to the wall-collapsing case (R= 13.3 {\AA})
by using the Landauer's formula \cite{Datta,Liu00}.
We chose the segment of the nanotorus containing 12 sections
that includes the wall-collapsing in its center as our sample
and connected it to ideal (5,5) SWNT leads. We obtained a
conductance of 0.5 $G_0$ ($G_{0}$ = $2e^2/h$) at the Fermi energy, a factor
of 4 reduction from the ideal straight (5,5) nanotube, consistent
with previous calculations \cite{Nardelli99}. This is, however, 
in dramatic contrast to the case of a local deformation, where a
small bending angle induces a two orders of magnitude reduction 
\cite{Liu00,Dai00}. We examined this issue by referring to the half width of 
the pronounced 
peak in the $\sigma$ bond charge (Fig. 5) and that of the deep valley in the 
$\pi$ bond charge (Fig. 6) for the 680-atom torus. It can be seen that 
both are about 0.24 rad, indicating that the transition from the $sp^2$- 
to the $sp^3$- bonding configuration only extends to a region of about 3{\AA}. It 
should be noted that the transition from the $sp^2$- to $sp^3$-bonding 
configuration induced by the manipulation of an AFM tip extends to a 
region of $\sim$ 12{\AA} at a small bending angle of 15$^o$ (see Fig. 3 in Ref. 8). 
The comparison explains why the reduction in conductance is relatively 
small even at large bending angles 
under a non-local distortion while a two-orders of magnitude reduction in 
conductance is observed under a local deformation.

The present study contrasts our previous studies \cite{Liu00,Dai00}
on the effects of local deformations on a SWNT. In particular, it brings out the 
important point that the deciding factor for the dramatic reduction of the 
conductance of metallic SWNTs is the combined effect of the transition from 
the $sp^2$- to $sp^3$-bonding configuration and the extent of the bending 
region where the transition occurs. With this study, we have now established 
a complete picture of the interplay between the mechanical
deformation, be it a non-local deformation or a local deformation, and the 
electronic properties of SWNTs. Our findings should provide useful guidelines 
for the design of nanoscale electromechanical devices based on SWNTs.

We acknowledge the support of the NSF grant (DMR-9802274) 
and the use of computing resources at the University of Kentucky Center for
Computational Sciences. We thank Prof. E. Kaxiras for his useful comments.

\newpage
\noindent{\bf FIGURE CAPTIONS}
\vskip 0.1 in
\noindent{Fig. 1: Four different stages of a nanotorus under a
 non-local deformation.}
 \vskip 0.1 in
\noindent{Fig. 2: The average strain energy per atom vs $\theta$.}
\vskip 0.1 in
\noindent{Fig. 3: The average bond length vs $\theta$. }
\vskip 0.1 in
\noindent{Fig. 4: The average coordination number vs $\theta$.}
\vskip 0.1 in
\noindent{Fig. 5: The average $\sigma$ bond charge vs $\theta$.}
\vskip 0.1 in
\noindent{Fig. 6: The average $\pi$ bond charge vs $\theta$.}
 

\begin{references}

\bibitem{Dekker98} A. Bezryadin, {\it et al.}, 
Phys. Rev. Lett. {\bf 80}, 4036 (1998).

\bibitem{Paulson99} S. Paulson {\it et al.}, 
Appl. Phys. Lett. {\bf 75}, 2936 (1999).

\bibitem{Crespi97}  V. Crespi, {\it et al.}, 
Phys. Rev. Lett. {\bf 79}, 2093 (1997).

\bibitem{Kane97} C.L. Kane and E.J. Mele, 
Phys. Rev. Lett. {\bf 78}, 1932 (1997).

\bibitem{Nardelli98} M. Nardelli, {\it et al.},  
Phys. Rev. Lett. {\bf 81}, 4656 (1998).

\bibitem{Avouris99}  A. Rochefort, {\it et al.}, 
xxx.lanl.gov/cond-mat/9904083 (1999).

\bibitem{Bernholc96} S. Iijima, {\it et al.},
J. Chem. Phys. {\bf 104}, 2089 (1996).

\bibitem{Liu00} L. Liu, {\it et al.}, 
Phys. Rev. Lett. {\bf 84}, 4950 (2000).

\bibitem{Dai00} T. Tombler, {\it et al.}, 
Nature {\bf 405}, 769 (2000).

\bibitem{Nardelli99} M. Nardelli and J. Bernholc, 
Phys. Rev. B {\bf 60}, 16338 (1999).

\bibitem{Avouris98} A. Rochefort, {\it et al.}, 
Chem. Phys. Lett. {\bf 297}, 45 (1998).

\bibitem{Avouris99a} R. Martel, {\it et al.},
Nature {\bf 398}, 299 (1999).

\bibitem{Avouris00} H.R. Shea, {\it et al.},
Phys. Rev. Lett. {\bf 84}, 4441 (2000).

\bibitem{Wu98} C.S. Jayanthi, {\it et al.}, 
 Phys. Rev. B {\bf 57}, 3799 (1998)

\bibitem{Menon93} M. Menon, {\it et al.},
Phys. Rev. B {\bf 48}, 8398 (1993).

\bibitem{Wu99} D. Alfonso, {\it et al.},
Phys. Rev. B {\bf 59}, 7745 (1999).

\bibitem{Datta} S. Datta {\it Electronic Transport in Mesoscopic
Systems} (Cambridge University Press, Cambridge, 1995).

\end{references}
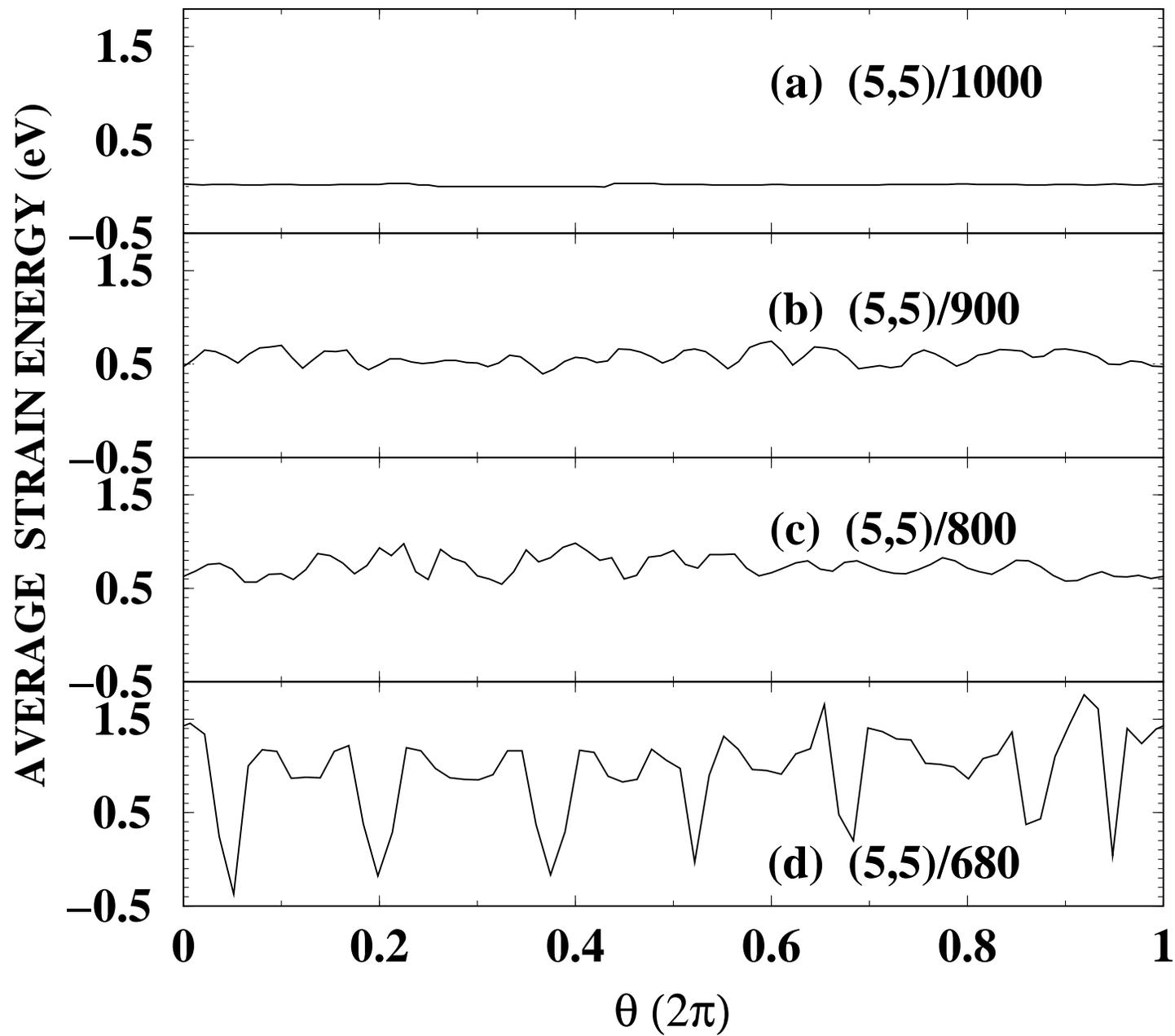
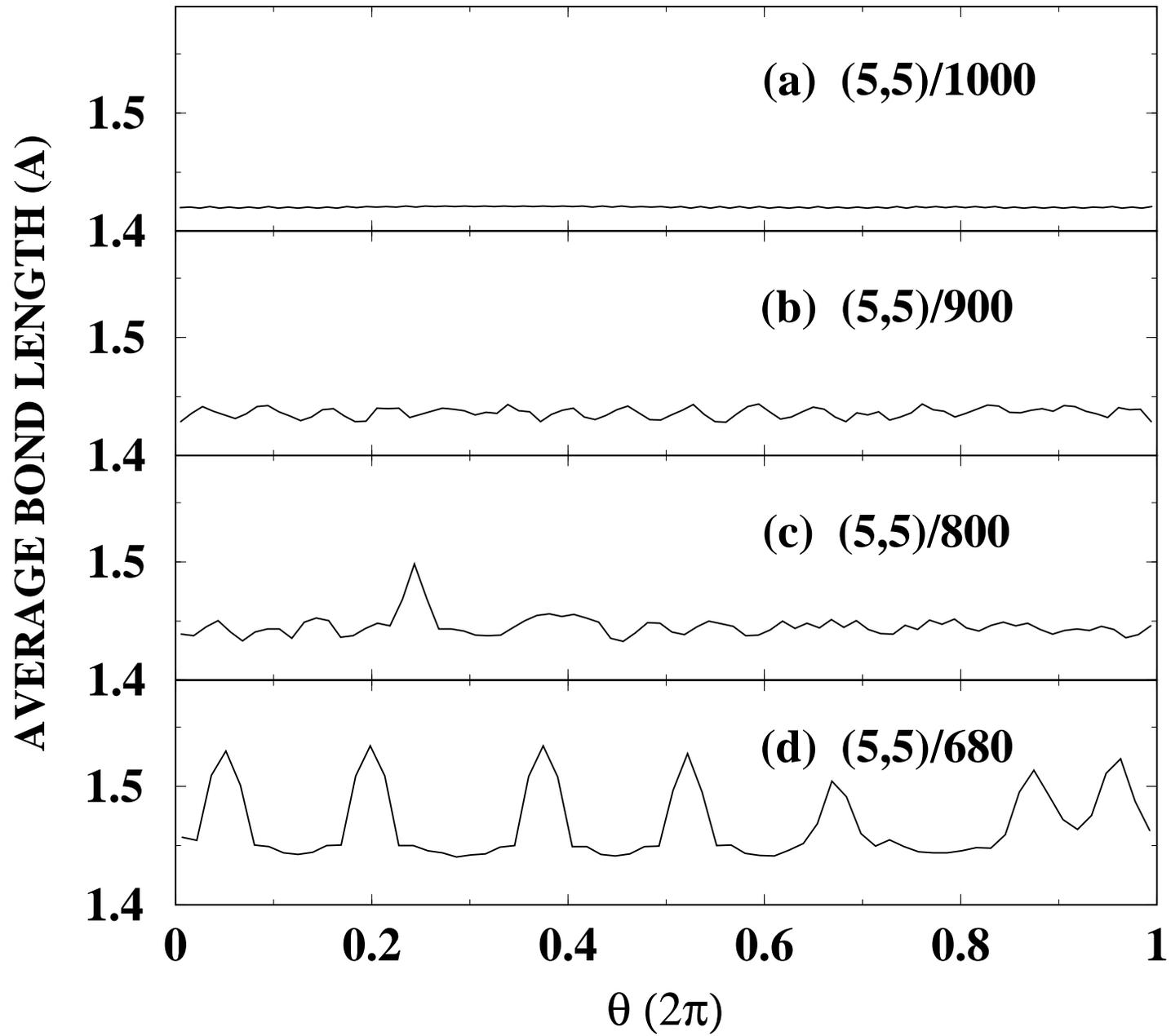
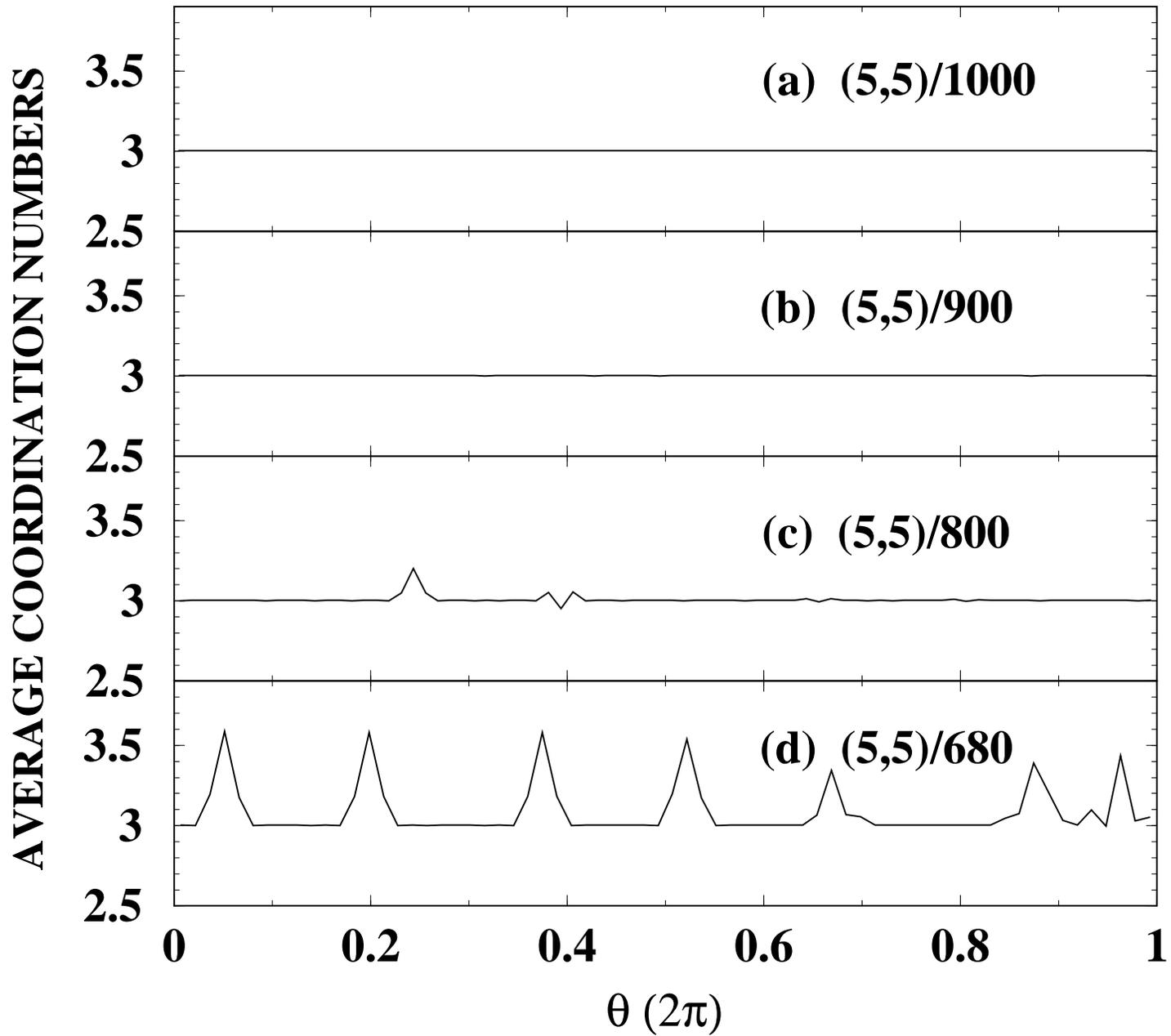
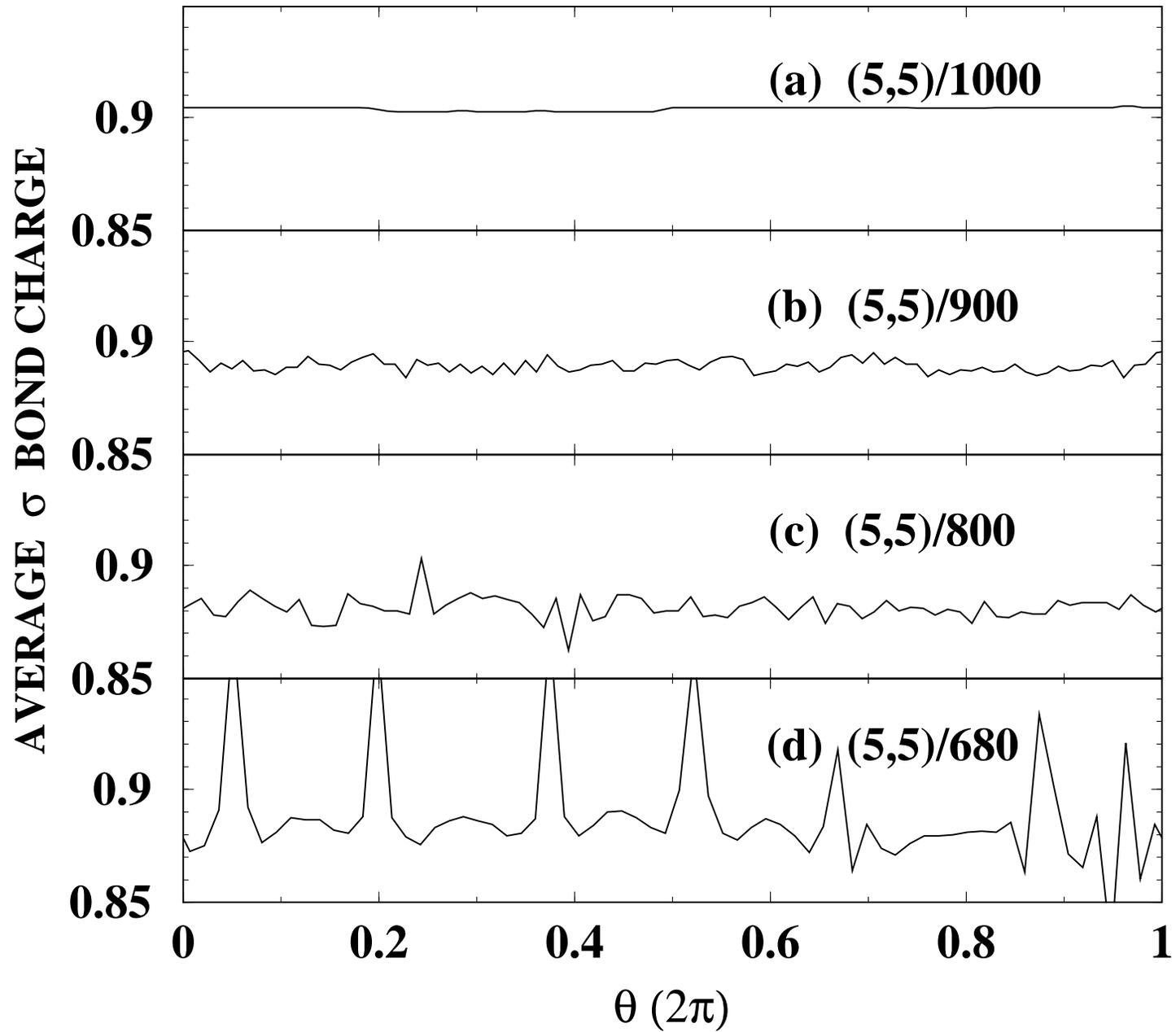
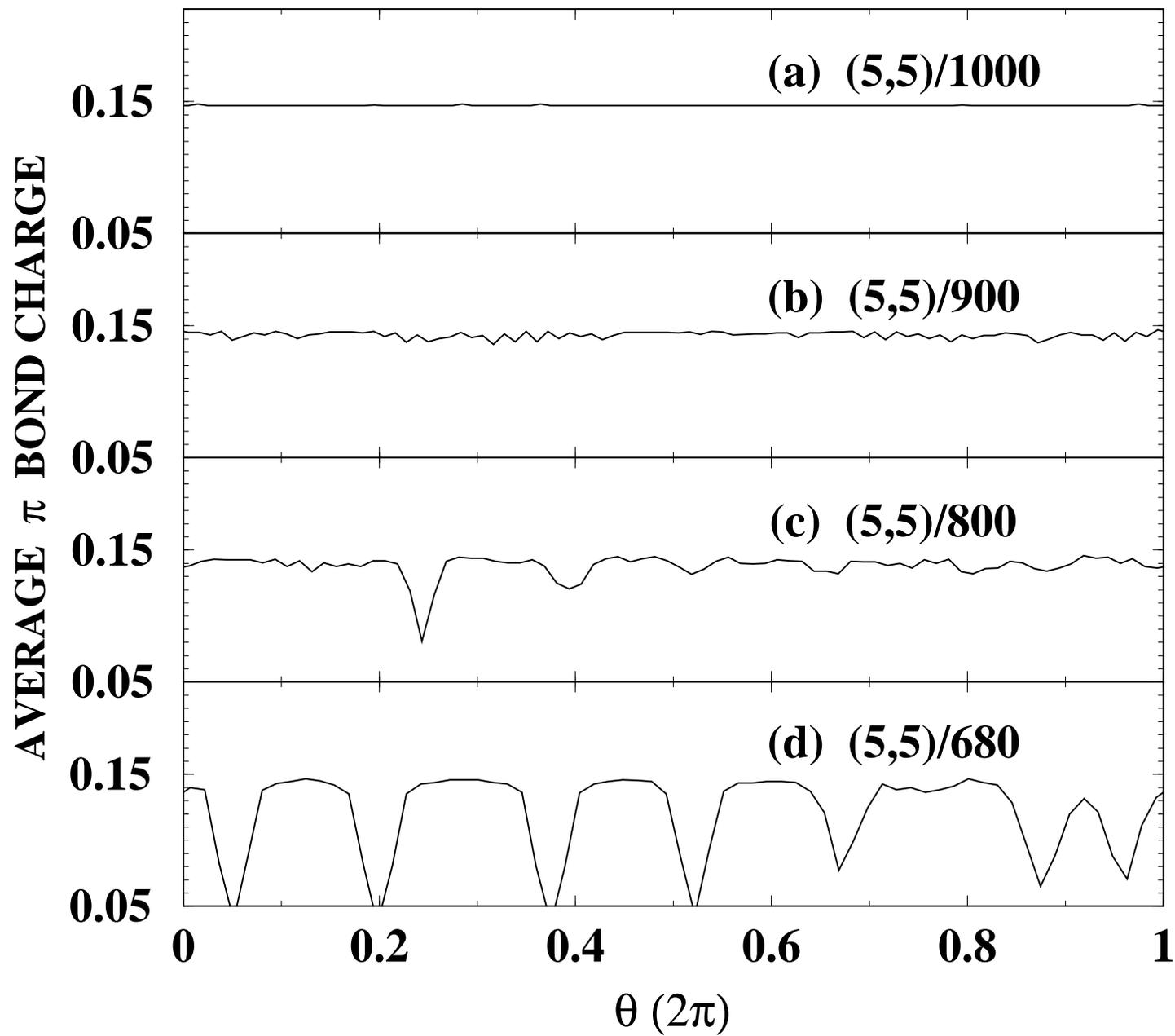
\end{document}